\documentclass[11pt,reqno,final]{article}
\usepackage[authoryear]{natbib}

\usepackage{setspace}
\usepackage{geometry}
\usepackage{amsmath,amsfonts,amssymb,amsthm,version}
\usepackage{mathrsfs,fancybox,pifont}
\usepackage{graphicx}
\usepackage{url,hyperref}
\usepackage[capitalize,nameinlink]{cleveref}
\usepackage[notcite,notref]{showkeys}
\usepackage{color}
\usepackage{subfigure,multirow}
\usepackage{epstopdf}
\usepackage{cases}
\usepackage{mathtools}
\usepackage{algorithm,algorithmic}
\usepackage{authblk}
\usepackage{enumerate}
\usepackage{multirow} 
\usepackage{booktabs}
\usepackage{lipsum}
\graphicspath{{./figures/}}

\allowdisplaybreaks[4]

\numberwithin{equation}{section}
\numberwithin{figure}{section}
\numberwithin{table}{section}

\theoremstyle{plain}
\newtheorem{theorem}{Theorem}[section]

\newtheorem{corollary}{Corollary}[section]

\theoremstyle{definition}

\newtheorem{assumption}{Assumption}[section]

\theoremstyle{remark}
\newtheorem{remark}{Remark}[section]

\newcommand{\norm}[1]{\left\lVert#1\right\rVert}

\title{Model Averaging under Flexible Loss Functions}
\author[a,b]{Dieqi Gu}
\author[c]{Qingfeng Liu}
\author[a,d]{Xinyu Zhang}
\affil[a]{International Institute of Finance, School of Management, University of Science and Technology of China, Hefei, China}
\affil[b]{Department of Management Sciences, College of Business, City University of Hong Kong, Kowloon, Hong Kong}
\affil[c]{Department of Industrial and Systems Engineering, Hosei University, Koganei, Tokyo, Japan}
\affil[d]{Academy of Mathematics and Systems Science, Chinese Academy of Sciences, Beijing, China}

\date{\today}

\begin{document}
\maketitle
	
\begin{abstract}
To address model uncertainty under flexible loss functions in prediction problems, we propose a model averaging method that accommodates various loss functions,
 including asymmetric linear and quadratic loss functions, as well as many other asymmetric/symmetric loss functions as special cases. The flexible loss function allows the proposed method to average a large range of models, such as the quantile and expectile regression models. To determine the weights of the candidate models, we establish a J-fold cross-validation criterion. Asymptotic optimality and weights convergence are proved for the proposed method. Simulations and an empirical application show the superior performance of the proposed method, compared with other methods of model selection and averaging.

		
\medskip
\noindent{\bf Supplemental Material:}: The software that supports the ﬁndings of this study is available within the paper and its Supplemental Information (
\url{https://pubsonline.informs.org/doi/suppl/10.1287/ijoc.2023.0291}
) as well as from the IJOC GitHub software repository (\url{https://github.com/INFORMSJoC/2023.0291}). The complete IJOC Software and Data Repository is available at \url{https://informsjoc.github.io/}.
\end{abstract}

\section{Introduction}
\label{sec:introduction}
Model uncertainty remains a major challenge in empirical and modeling efforts. 
In practice, we may encounter many different models in a prediction problem. However, determining which model to use for predicting the response variable can be challenging, leading to the model uncertainty problem.   In many cases, models differ in terms of the explanatory variables they include. At this point, model uncertainty actually comprises variable uncertainty. To put it another way, we are often confronted with numerous candidate models including different explanatory variables, which leads to uncertainty regarding which model and variables we should use. 
Intuitively, one popular approach for dealing with this problem is variable selection or, more generally, model selection; see, for example, \cite{tibshirani1996regression}, \cite{akaike1998information}, 
\cite{zucchini2000introduction}, \cite{ding2018model} and \cite{zhang2023splicing}.

Meanwhile, as an alternative to model selection, model averaging has attractive properties that help address model uncertainty from a different perspective. Specifically, rather than selecting an individual model, model averaging combines all candidate models and averages their estimators or predictors using certain weights. 
Unlike model selection, which ``puts all our inferential eggs in one unevenly woven basket" \citep{Longford2005editorial},  model averaging mitigates prediction errors because it reveals useful information from the form of the relationship between the response and explanatory variables and provides a type of insurance against a single selected model exhibiting a poor fit \citep{bates1969combination, leung2006information}. Additionally, it is often the case that several models fit the data equally well, but may diﬀer substantially in terms of the variables included and may lead to diﬀerent predictions \citep{miller2002subset}; thus, combining these models seems to be more reasonable than choosing one of them. Consequently, model averaging methods have the potential to overperform model selection.
Model averaging has been extensively studied over the past two decades; see, for example, \cite{hansen2007least}, \cite{claeskens2008model}, \cite{hansen2012jackknife}, 
\cite{liu2013heteroscedasticity}, \cite{nelson2021reducing} and \cite{li2024partial}.
Due to its good predictive ability, model averaging can be used to address prediction problems in all areas of science, including management, ﬁnance, economics, biology, etc. More examples of its application can be found in the research of \cite{zhang2019demand} and \cite{steel2020model}.

In addition to the challenge of model uncertainty, we are often concerned with the loss function of estimation in specific prediction problems, as it can profoundly affect our prediction.
 In many prediction problems, the most commonly used loss functions---such as the mean squared error and mean absolute error---are symmetric. Symmetric loss functions are used assuming that upward and downward biases are equally important.
However, in some cases, preferences regarding the costs of positive and negative prediction errors may differ; thus, we require a flexible loss function that allows for asymmetries.
Suppose that we study the side effects of a drug, while the downsides from overestimating or underestimating the side effects of the drug differ significantly. Overestimating the side effects may lead to a decrease in the dose and, thus, a longer treatment time. However, underestimating the drug’s side effects may lead to an increase in the dose, thereby threatening the patient's life.
Similarly, in macroeconomic forecasting, optimistic GDP-growth forecasts can lead to lower projections of nominal deﬁcit, thus leaving some leeway for the required fiscal adjustment (see \citealt{christodoulakis2008assessment}). 
Some studies illustrate this phenomenon, such as those conducted by  \cite{ray2006asymmetric} and \cite{chen2021asymmetric}.
Therefore, in situations where preferences for the costs of positive and negative prediction errors vary---as in the abovementioned medical or financial contexts---an asymmetric loss function is preferred.

Specifically, we aim to estimate the response variable $y$ under a particular loss function $L(\cdot)$, conditioned on the information provided by observed related covariates $\mathbf{x}\in \mathbb{R}^k$.
In practice, when historical observations $\{(\mathbf{x}_i , y_i )\}, i=1,\ldots, n$ are available, we aim to solve the following optimization problem: 
$$
 \min_{\boldsymbol{\theta} \in \mathbb{R}^k}
 \sum_{i=1}^{n}L\left(y_{i}-\boldsymbol{\theta}^{\prime} \mathbf{x}_{i}\right),
$$
where $L(\cdot)$ is any kind of loss function and may be symmetric or asymmetric. We can handle different problems with different selections of $L(\cdot)$.
Some theories regarding estimation under general loss functions have been presented by \cite{christoffersen1996further}, \cite{christoffersen1997optimal}, and \cite{granger1999outline}, among others.


Interestingly,  \cite{elliott2005estimation} proposed a flexible loss function  that can help us handle asymmetric and symmetric losses in a common framework. 
Following \cite{elliott2005estimation}, we define the loss function as
\begin{equation}
\label{eq:lossfun}
\rho_{\tau, p}(\lambda)=|\tau-\mathbf{1}\{\lambda \leq 0\}| \cdot
|\lambda|^{p}, 	
\end{equation}
where $\tau \in(0,  1)$ represents  the asymmetry parameter,  while $p = 1, 2$, and $\mathbf{1}\{\cdot \}$ denotes the indicator function. Note that for $p=1$ and $\tau\neq 0.5$,  Equation \eqref{eq:lossfun} is an asymmetric linear loss function, whereas for $p=2$ and $\tau\neq 0.5$, Equation \eqref{eq:lossfun} is an asymmetric quadratic loss function. Additionally, if $\tau=0.5$,  then Equation \eqref{eq:lossfun} is symmetric.


Some studies have explored model averaging under 
asymmetric loss functions. For instance, 
\cite{lu2015jackknife} proposed a quantile regression (QR) model averaging method based on the jackknife criterion and used an asymmetric linear loss function. \cite{wang2023jackknife} extended this to the high-dimensional case. \cite{guo2022model} considered QR model averaging under varying coefficient models based on the jackknife criterion. Finally, \cite{xumodel} considered model averaging under functional linear QR models via J-fold
cross-validation (JCV).
Further, \cite{tu2020jackknife} proposed an expectile regression (ER) model averaging method based on the jackknife criterion using an asymmetric quadratic loss function. Meanwhile, \cite{bai2022optimal} considered JCV ER model averaging.

In the present study, we investigate model averaging under the 
flexible loss function \eqref{eq:lossfun}. This function enables us to propose a model averaging method that can be implemented in various situations and types of models---e.g., QR, ER, and linear models with either asymmetric or symmetric loss. Importantly, the flexible loss function helps us obtain common theoretical results of model averaging for the aforementioned types of models.

To implement the proposed method, we must first determine the weights of model averaging. Methods based on cross-validation (CV) criteria are widely used in model averaging because they can approximate the prediction loss well and are easy to implement. For instance, \cite{hansen2012jackknife} proposed jackknife model averaging for least squares regression, while \cite{zhang2013model} extended it to dependent data. \cite{cheng2015forecasting} extended the jackknife criterion to leave-$h$-out CV criteria for prediction, in combination with factor-augmented regression. \cite{gao2016model} extended this approach to leave-subject-out CV under a longitudinal data setting. Recently, several studies have used JCV to select model averaging weights, such as \cite{zhang2018functional} and \cite{liu2019detectinga}. 
The basis of JCV is dividing the data set into J groups and treating each group as a validation data set to evaluate the model.
JCV can help in approximating the prediction error well. Moreover,
compared to the jackknife criterion, JCV is much less computationally expensive because we must only compute the estimates J times.
Hence, we adopt a method based on the JCV criterion to determine the weights of our model averaging method (hereafter, JCV model averaging [JCVMA]).

We investigate the asymptotic properties of the proposed JCVMA method and find some notable theoretical results. If all candidate models are misspecified, JCVMA is asymptotically optimal in terms of minimizing the excess final prediction error proposed by \cite{xumodel}. The excess final prediction error, which removes the random error term of the final prediction error, is a more appropriate measure of asymptotic optimality. Therefore, our findings improve the asymptotic optimality based on the final prediction error used in the aforementioned  studies on model averaging under asymmetric loss functions. 
We also consider a situation in which at least one correct model is included in the candidate models,  which is rarely considered in the related literature.
We demonstrate that our model averaging estimator of coefficients is consistent. Additionally, we prove that the sum of the weights assigned to the correct models converges to 1 in probability.  To the best of our knowledge, this property has never been demonstrated in previous studies on model averaging under asymmetric loss.  
Monte Carlo simulations indicate the superiority of the proposed method. We apply our method to predict the collection volume of shipping carrier stores under a flexible loss function, thereby demonstrating its advantages.

The main contributions of our work, compared with previous studies, are as follows.
First, we adopt a flexible loss function so that JCVMA can handle various types of models with either asymmetric or symmetric losses. Consequently, our asymptotic results are generally applicable to models with both types of loss functions.
Second, we derive the asymptotic results under the condition that all candidate models are misspecified and that there exist correct models. This highlights the performance of our method in both cases.
Third, all theoretical results are established under the premise that the numbers of covariates and candidate models can diverge as the sample size increases.
Fourth, our asymptotic optimality is based on the excess final prediction error, which is more appropriate than that used in most previous studies.

The remainder of this paper is organized as follows. Section \ref{sec:MA} describes our model averaging estimation method and proposes the JCV criterion.
Section \ref{sec:Performance_Guarantees} develops some asymptotic properties and conducts simulation studies to illustrate the finite sample performance. Section \ref{sec:emprical} applies the proposed method to predict the collection volume of the shipping carrier stores. Finally, Section \ref{sec:conclusions} presents the conclusions of this study. Proofs of the main results and some experimental results are provided in the Supplementary Materials \citep{gu2024model}.

\section{Methodology}
\label{sec:MA}
In this section, we present our model and methodology. We explore model averaging under the flexible loss function \eqref{eq:lossfun}, which enables us to incorporate a wide range of models, including QR and ER models. Additionally, we establish a J-fold cross-validation criterion to determine the weights of the candidate models.

\subsection{Model Averaging under Flexible Loss }
Let $\left\{\left(y_{i},  \mathbf{x}_{i}\right)\right\}_{i=1}^{n}$ be a group of independent and identically distributed (IID) random samples, where $y_i$ is a scalar dependent variable and $\mathbf{x}_{i}=\left(x_{i 1},  x_{i 2},  \ldots\right)$ is an explanatory variable of countably infinite dimension. Without loss of generality, we assume that $x_{i1}=1$. Following  \cite{koenker1982robust}, we consider the following data generation process (DGP):
\begin{equation}
\label{eq:DGP}
y_{i}=\sum_{j=1}^{\infty} \beta_{j} x_{i j}+\left(\sum_{j=1}^{\infty} \alpha_{j} x_{i j}\right) \epsilon_{i}, 
\end{equation}
where $\alpha_j $ and $\beta_j$ are unknown parameters, $\alpha_1=1$ for identifiability, and $\epsilon_{i}$ is an IID unobservable error term independent of $\mathbf{x}_{i}$. This infinite-dimensional linear model allows heteroskedasticity and is applicable to many situations.

We use the flexible loss function \eqref{eq:lossfun}.
Define
$\mathscr{E}_{\tau, p}(\epsilon_i )= \arg \min _{q \in \mathbb{R}} \operatorname{E}\rho_{\tau, p}\left(\epsilon_{i}-q
 \right) = \mathscr{E}_{\tau, p} $ and
rewrite the model \eqref{eq:DGP} as follows:
\begin{align}
\label{eq:model}
y_{i}=&\sum_{j=1}^{\infty}\left(\beta_{j}+\alpha_{j} \mathscr{E}_{\tau, p}\right) x_{i j}+\left(\sum_{j=1}^{\infty} \alpha_{j} x_{i j}\right)\left(\epsilon_{i}-\mathscr{E}_{\tau, p}\right) \notag \\= & \sum_{j=1}^{\infty} \theta_{j} x_{i j}+\varepsilon_{i}\notag \\
= & \mu_i +\varepsilon_i ,  
\end{align}
where $\theta_{j}=\beta_{j}+\alpha_{j} \mathscr{E}_{\tau, p}$, $\varepsilon_{i}=\left(\sum_{j=1}^{\infty} \alpha_{j} x_{i j}\right)\left(\epsilon_{i}-\mathscr{E}_{\tau, p}\right)$ satisfies the following:
\begin{equation}
\label{eq:generalloss}
\arg \min _{q \in \mathbb{R}} \operatorname{E}\left[ \rho_{\tau, p}\left(\varepsilon_{i}-q
 \right)\mid \mathbf{x}_i\right]=0, 
\end{equation}
and $\mu_i=\sum_{j=1}^{\infty} \theta_{j} x_{i j}$\footnote{We frequently suppress the dependence of $\theta_{j}, \varepsilon_{i}, \mu_i$ on $\tau, p$ for notational simplicity.}. 
Thus, $$ \arg \min _{q \in \mathbb{R}} \operatorname{E}\left[ \rho_{\tau, p}\left(y_{i}-q
 \right)\mid \mathbf{x}_i\right]=\mu_i.$$

In most articles that explore asymmetric loss, the parameters $p$ and $\tau$ are typically provided directly (e.g., \citealt{christoffersen1996further}, \citealt{demetrescu2019predictive},  \citealt{salari2022realtime}). 
In general, we determine our parameters $p$ and $\tau$ based on the nature of the problem being addressed, considering the perspectives and requirements of the forecasters. Specifically, the parameter $p=1$ is preferable when dealing with outliers whose influence should be reduced. The parameter $p=2$ is suitable when outliers significantly impact model performance, as seen in financial or security domains where a larger penalty for errors is warranted.
The parameter $\tau$ represents the asymmetry parameter, which depends on the degree of loss asymmetry and is influenced by forecasters' preferences and knowledge.
In the example of studying a drug’s effects, since underestimation has worse consequences than overestimation, researchers may opt for $\tau>0.5$.  In the example of GDP-growth forecasting, as optimistic forecasting can leave some leeway for the required fiscal adjustment,  $\tau>0.5$ might be chosen \citep{christodoulakis2008assessment}. 

Some studies delve into determining the asymmetry parameter $\tau$ when having observed a sequence of forecasts; see \cite{elliott2005estimation} and \cite{tsuchiya2016assessing}.
Specifically, $\tau$ can be estimated based on the moment conditions for a given $p$ with instrumental variables. Researchers have employed direct estimation methods  \citep{elliott2005estimation} or utilized generalized methods of moments  to estimate $\tau$
\citep{tsuchiya2016assessing}. Particularly,   \cite{tsuchiya2016assessing} estimated the asymmetry parameter $\tau$ in the forecasts of the real GDP growth of the next fiscal year in Japan and discovered that the government employs $\tau=0.7$, while the International Monetary Fund adopts $\tau=0.65$.

We consider $M$ approximate linear models containing different covariates to approximate the true model \eqref{eq:model}, where the set of candidate models can be nested or non-nested, while $M$ can go to infinity with sample size $n$. For $m=1,\ldots, M$, let $k_m$ be  the number of covariates in the $m$th  model. The $m$th candidate model is written as follows:
$$
y_{i}=\boldsymbol{\theta}_{(m)}^{\prime} \mathbf{x}_{i(m)}+\varepsilon_{i(m)}=\sum_{j=1}^{k_{m}} \theta_{j(m)} x_{i j(m)}+\varepsilon_{i(m)}, 
$$
where $\boldsymbol{\theta}_{(m)} =\left(\theta_{1(m)},  \ldots,  \theta_{k_{m}(m)}\right)^{\prime},  \mathbf{x}_{i(m)}=\left(x_{i 1(m)},  \ldots,  x_{i k_{m}(m)}\right)^{\prime}$,  $x_{i j(m)},  j=1,  \ldots,  k_{m}$ are variables in $\mathbf{x}_{i}$ that appear as regressors in the $m$th model, $\theta_{j(m)}$ is the corresponding coefficient, and $\varepsilon_{i(m)}=\mu_{i}-\sum_{j=1}^{k_{m}} \theta_{j(m)} x_{i j(m)}+\varepsilon_{i}$ stands for the total error in the $m$th model. Specifically, $k_m$ is allowed to diverge with $n$.

For the $m$th candidate model,  the estimator of $\boldsymbol{\theta}_{(m)}$ is given by:
\begin{equation}
\label{eq:theta}
\begin{aligned}
\hat{\boldsymbol{\theta}}_{(m)} 
&=\arg \min _{\boldsymbol{\theta}_{(m)}} \sum_{i=1}^{n} \rho_{\tau, p}\left(y_{i}-\boldsymbol{\theta}_{(m)}^{\prime} \mathbf{x}_{i(m)}\right).
\end{aligned}
\end{equation}
Note that when $p=1$, \eqref{eq:theta} is a QR estimator \citep{koenker1978regression}.
When $p=2$, \eqref{eq:theta} is an ER estimator \citep{newey1987asymmetric}.
Moreover, when $p=2$ and $\tau=0.5$, \eqref{eq:lossfun} is the traditional squared loss function and \eqref{eq:theta} reduces to the ordinary least squares (OLS) estimator. When $p=1$ and  $\tau=0.5$, \eqref{eq:lossfun} is the absolute loss function and \eqref{eq:theta} reduces to the minimum median estimator.

Let $\mathbf{w} = \left(w_{1},  \ldots,  w_{M}\right)^{\prime}$ be the weight vector in $\mathcal{W} =\left\{\mathbf{w} \in[0, 1]^{M}: \sum_{m=1}^{M} w_{m}=1\right\}$. 
For $i=1,  \ldots,  n$, the model averaging estimator of $\mu_{i}$ is given by:
\begin{equation}
\label{eq:mu}
\hat{\mu}_{i}(\mathbf{w})=\sum_{m=1}^{M} w_{m} \mathbf{x}_{i(m)}^{\prime} \hat{\boldsymbol{\theta}}_{(m)}.
\end{equation}

\subsection{Weight Choice Criterion}
We use JCV to choose the weights. 
In the case of the OLS estimator, the jackknife criterion is typically used because the parameter estimators under leave-one-out CV (LOOCV) can be calculated based on the original parameter estimators. Therefore, there is no need to calculate $n$ times and the computational complexity substantially decreases.
However, under flexible loss, there is no simple calculation method for the parameter estimator under LOOCV. If $n$ is large, using LOOCV is computationally expensive. Therefore, we choose to use JCV to select the weights, as it is much less computationally expensive.

To implement JCV, we randomly divide the dataset into $J$ groups ($J \geq 2$ is a positive finite integer), each with $Q=n/J$ observations.
When $n/J$ is not an integer, let each of the first $J-1$ groups contain $Q=\lfloor n/J \rfloor$ observations and the last group contain $n-(J -1)Q$ observations, where $\lfloor \cdot \rfloor$ denotes the integer part of $\cdot$. 
To simplify the proof, let $n/J$ be an integer. 

For $m = 1, \ldots , M$, let $\tilde{\boldsymbol{\theta}}_{(m)}^{[-j]}$ denote
the estimator of  $\boldsymbol{\theta}_{(m)}$ in the $m$th model after excluding the $j$th group of observations. For $i=(j-1)Q+1,\ldots, jQ$ in the $j$th group, we form the estimator of $\mu_i$ as follows:
$\tilde{\mu}_{i}^{[-j]}(\mathbf{w}) = \sum_{m=1}^{M} w_{m} \mathbf{x}_{i(m)}^{ \prime} \tilde{\boldsymbol{\theta}}_{(m)}^{[-j]}.$ 
Furthermore, we define $\mu_i^*(\mathbf{w})= \sum_{m=1}^{M} w_{m} \mathbf{x}_{i(m)}^{\prime} {\boldsymbol{\theta}}_{(m)}^*,  i=1, 2, \ldots , n.$ 
Our JCV criterion is formulated as follows:
\begin{equation}\label{eq:weight_criterion}
\operatorname{CV}_{n}^J(\mathbf{w})=\frac{1}{n} 
\sum_{j=1}^{J} \sum_{q=1}^{Q}
\rho_{\tau, p}\left(y_{(j-1)Q+q}-\tilde{\mu}_{(j-1)Q+q}^{[-j]}
(\mathbf{w})\right).
\end{equation}
The JCV weight vector $\hat{\mathbf{w}}=\left(\hat{w}_{1},  \ldots,  \hat{w}_{M}\right)'$ is obtained by choosing $\mathbf{w} \in \mathcal{W}$ to minimize the above criterion function; that is,
\begin{equation}
\label{eq:argmin}
\hat{\mathbf{w}}=\arg \min _{\mathbf{w} \in \mathcal{W}} \operatorname{C V}^J_{n}(\mathbf{w}).
\end{equation}
Then, by substituting $\hat{\mathbf{w}}$ into \eqref{eq:mu}, we obtain the JCVMA estimator as follows: $$
\hat{\mu}_{i}(\hat{\mathbf{w}})=\sum_{m=1}^{M} \hat{w}_{m} \mathbf{x}_{i(m)}^{\prime} \hat{\boldsymbol{\theta}}_{(m)}.
$$

According to \cite{lu2015jackknife} and \cite{bai2022optimal},
we can convert the constrained optimization problem \eqref{eq:argmin} to the following linear or quadratic programming problem, which can be solved quickly and stably.
When $p=1$, \eqref{eq:argmin} is equivalent to a linear programming problem of the following form:
$$
\begin{aligned}
&\min _{\mathbf{w},  \mathbf{u},  \mathbf{v}}\left\{
\tau \mathbf{1}_{n}^{\prime} \mathbf{u}+(1-\tau) \mathbf{1}_{n}^{\prime} \mathbf{v}\right\} \\
\text { s.t. } & \sum_{m=1}^{M} w_{m} \mathbf{x}_{((j-1)Q+q)(m)}^{ \prime} \tilde{\boldsymbol{\theta}}_{(m)}^{[-j]}
+u_{(j-1)Q+q}-v_{(j-1)Q+q}=y_{(j-1)Q+q},  \\
&u_{(j-1)Q+q}\geq 0,  v_{(j-1)Q+q}\geq 0 ,  j=1,  \ldots,  J,  q=1,  \ldots,  Q,   \\
&\sum_{m=1}^{M} w_{m}=1, 
0 \leq w_{m} \leq 1,  m=1,  \ldots,  M, 
\end{aligned}
$$
where $\mathbf{u} =\left(u_{1},  u_{2},  \ldots,  u_{n}\right)^{\prime}$ and  $\mathbf{v} =\left(v_{1},  v_{2},  \ldots,  v_{n}\right)^{\prime}$ are the positive and negative slack variables, respectively, and $\mathbf{1}_{n}$ is a $n \times 1$ vector of ones.
When $p=2$, \eqref{eq:argmin} is equivalent to a quadratic programming problem of the form
$$
\begin{aligned}
&\min _{\mathbf{w},  \mathbf{u},  \mathbf{v}}\left\{
\tau \mathbf{u}^{\prime} \mathbf{u}+(1-\tau) \mathbf{v}^{\prime} \mathbf{v}\right\} \\
\text { s.t.} & \sum_{m=1}^{M} w_{m} \mathbf{x}_{((j-1)Q+q)(m)}^{ \prime} \tilde{\boldsymbol{\theta}}_{(m)}^{[-j]}
+u_{(j-1)Q+q}-v_{(j-1)Q+q}=y_{(j-1)Q+q},  \\
&u_{(j-1)Q+q}\geq 0,  v_{(j-1)Q+q}\geq 0 ,  j=1,  \ldots,  J,  q=1,  \ldots,  Q,   \\
&\sum_{m=1}^{M} w_{m}=1, 
0 \leq w_{m} \leq 1,  m=1,  \ldots,  M, 
\end{aligned}
$$
where $\mathbf{u} =\left(u_{1},  u_{2},  \ldots,  u_{n}\right)^{\prime}$ and $\mathbf{v} =\left(v_{1},  v_{2},  \ldots,  v_{n}\right)^{\prime}$ are the positive and negative slack variables, respectively.

\section{Performance Guarantees}
\label{sec:Performance_Guarantees}
Following \cite{zhang2020parsimonious}, \cite{racine2022optimal}, and \cite{liu2022frequentist},
we call a model that includes all regressors with nonzero coefficients and only these regressors a \textit{true model}. This is also referred to as the \textit{just-fitted model} \citep{zhang2019inference}. 
A \textit{correct model} refers to a model in which the regressors contain all regressors in the true model. 

In this section, we provide theoretical performance guarantees for the proposed JCVMA method. 
In reality, we have no idea whether the true model or the correct models are included in the set of candidate models. Therefore, we consider two scenarios to cover all possible situations when investigating the asymptotic properties of the JCVMA method. The first scenario is where all candidate models are misspecified. The second scenario is where at least one of the candidate models is correct. These two scenarios provide us with deeper insight into the statistical properties of our method and lead to notable theoretical results.

In the first scenario, we show that our JCVMA-selected weight vector $\hat{\mathbf{w}}$ is asymptotically optimal for minimizing the excess final prediction error.
In the second scenario, we establish the estimation consistency of the proposed model averaging estimator. Additionally, we find that the sum of the weights assigned to the correct models converges to 1 in probability.

Then, we conduct Monte Carlo simulations to emulate both scenarios and evaluate the finite sample performance of the proposed JCVMA method, which supports our theoretical results.

\subsection{Notation}
We adopt the following notation. First, all limit processes are with respect to $n \to \infty$, unless otherwise specified. The notations $\stackrel{p}{\longrightarrow}$ and $\stackrel{d}{\longrightarrow}$ denote convergences in probability and  distribution, respectively. ``a.s.''  denotes ``almost surely."
For a $d \times 1$ vector $\mathbf{v}=\left(v_1, \ldots, v_d\right)^{\prime}$, we use $\norm{\cdot}$ to denote its Euclidean norm; that is, $\norm{\mathbf{v}}=\left(\sum_{i=1}^d v_i^2\right)^{1 / 2}$.
For an $m \times n$ matrix $A$, we denote its transpose by $A^{\prime} $.
When $A$ is symmetric, we use $\lambda_{ \max }(A)$ and $\lambda_{ \min }(A)$  to denote the largest and smallest eigenvalues, respectively.

\subsection{Asymptotic Optimality}
Here, we present an important result on the asymptotic optimality of the selected weight vector $\mathbf{w}$ for minimizing its excess final prediction error.

Let $(y_{n+1},  \mathbf{x}_{n+1})$ be an independent copy of $\left(y_{i},  \mathbf{x}_{i}\right), i=1, 2, \ldots,  n $. We define
$\hat{\mu}_{n+1}(\mathbf{w})= \sum_{m=1}^{M} w_{m} 
\mathbf{x}_{n+1(m)}^{\prime} \hat{\boldsymbol{\theta}}_{(m)},
$
where $\mathbf{x}_{n+1(m)} =(x_{n+1, 1(m)},  \ldots,$  $x_{n+1, k_{m}(m)})^{\prime}$ and $x_{n+1, 1(m)}, \ldots,  x_{n+1, k_m(m)} $ are variables in $\mathbf{x}_{n+1}$ that appear as regressors in the $m$th model for $m=1,\ldots, M $.
Given $\hat{\mathbf{w}}$, we can obtain the model averaging estimator of $\mu_{n+1}$ as
$$
\hat{\mu}_{n+1}(\hat{\mathbf{w}})= \sum_{m=1}^{M} \hat{w}_{m} 
\mathbf{x}_{n+1(m)}^{\prime} \hat{\boldsymbol{\theta}}_{(m)}.
$$

Let $\mathscr{D}_{n} =$ $\left\{\left(y_{i},  \mathbf{x}_{i}\right)\right\}_{i=1}^{n}$. 
Define 
the final prediction error (FPE) or out-of-sample prediction error used by \cite{lu2015jackknife} as
$$
\operatorname{FPE}_{n}(\mathbf{w})=
\operatorname{E}\left[\rho_{\tau, p}\left(y_{n+1}-
\hat{\mu}_{n+1}(\mathbf{w})
\right) \mid \mathscr{D}_{n}\right].
$$
According to \cite{xumodel}, we define the excess final prediction error  (EFPE) as 
\begin{align*}
\operatorname{EFPE}_{n}(\mathbf{w})&=\operatorname{FPE}_{n}(\mathbf{w})-\operatorname{E}\left[\rho_{\tau,p}\left(\varepsilon_{n+1}\right)\right]\\
&=\operatorname{E}\left[\rho_{\tau, p}\left(y_{n+1}-
\hat{\mu}_{n+1}(\mathbf{w})
\right) \mid \mathscr{D}_{n}\right]-\operatorname{E}\left[\rho_{\tau,p}\left(\varepsilon_{n+1}\right)\right].
\end{align*}
Clearly, $\operatorname{EFPE}_{n}(\mathbf{w})\geq0$ for any $\mathbf{w} \in \mathcal{W}$ (see Lemma 4 in  the Supplementary Materials), and when $\hat{\mu}_{n+1}(\mathbf{w}_0)=\mu_{n+1}$ a.s. for some $\mathbf{w}_0 \in \mathcal{W}$, $\operatorname{EFPE}_{n}(\mathbf{w}_0)=0$ a.s.

For the convenience of studying the candidate models under misspecification, we define the pseudo-true parameter as follows:
$$
\boldsymbol{\theta}_{(m)}^{*} = \arg \min _{\boldsymbol{\theta}_{(m)}} \operatorname{E}\left[\rho_{\tau, p}\left(y_{i}-\mathbf{x}_{i(m)}^{\prime} \boldsymbol{\theta}_{(m)}\right)\right] .
$$
As $\rho_{\tau, p}(\lambda)$ is convex, $\boldsymbol{\theta}_{(m)}^{*}$ exists and is unique.
Then, we define 
$\mu_{n+1}^*(\mathbf{w})= \sum_{m=1}^{M} w_{m} \mathbf{x}_{n+1(m)}^{\prime} {\boldsymbol{\theta}}_{(m)}^*
$ and
\begin{align*}
\begin{aligned}
\operatorname{EFPE}_{n}^{*}(\mathbf{w}) 
&=\operatorname{E}\left[\rho_{\tau,p}\left(y_{n+1}-\mu_{n+1}^{*}\left(\mathbf{w}\right)\right)\right]
-\operatorname{E}\left[\rho_{\tau,p}\left(\varepsilon_{n+1}\right)\right],\end{aligned}
\end{align*}
and let $\xi_{n}=\inf _{\mathbf{w} \in \mathcal{W}} \operatorname{EFPE}_{n}^{*}(\mathbf{w}) $ denote the minimum EFPE in the class of model averaging estimators associated with  $\boldsymbol{\theta}_{(m)}^*$.

We consider the following two situations:

\noindent
(i) all candidate models are misspecified, but $\xi_n \to 0$; and

\noindent
(ii) at least one correct model exists in the set of candidate models.

\noindent
Under either situation, we can show that $\inf_{\mathbf{w}\in \mathcal{W}} \operatorname{EFPE}_{n}(\mathbf{w}) \stackrel{p}{\longrightarrow}  0$ under some regularity conditions (see Lemma 4 in the Supplementary Materials). Hence,
compared with $\operatorname{EFPE}_{n}(\mathbf{w})$, $\operatorname{FPE}_{n}(\mathbf{w})$ has a nonnegative random error term $\operatorname{E}\left[\rho_{\tau,p}\left(\varepsilon_{n+1}\right)\right]$, which is a constant and, thus, the dominant term in the FPE. This means that the asymptotic optimality based on the FPE makes no sense.
Therefore,
we use the EFPE as a criterion to evaluate the prediction performance, rather than the FPE.  

Let $f\left(\cdot \mid \mathbf{x}_{i}\right)$ and $F\left(\cdot \mid \mathbf{x}_{i}\right)$ denote the conditional probability density function (PDF) and cumulative distribution function (CDF) of $\varepsilon_{i}$ given $\mathbf{x}_{i}$, respectively. We define $\psi_{\tau}(u)= \tau- \mathbf{1}\{u \leq 0\} $, $\Psi_{\tau}(u) =|\tau-\mathbf{1}\{u \leq 0\}| u$ and $\bar{k}=\max_{1\leq m \leq M}k_m. $
We now state some regularity assumptions for establishing the asymptotic properties.

\begin{assumption}
\label{assumption1}
$\operatorname{E}\left(\mu_{i}^{16}\right)<\infty$, $\operatorname{E}\left(\varepsilon _{i}^{16}\right)<\infty$, and $\sup _{j \geq 1} \operatorname{E}\left(x_{i j}^{16}\right) \leq c_{\mathbf{x}}$ for some $c_{\mathbf{x}}<\infty$.
\end{assumption}

\cref{assumption1} imposes restrictions on the moments. 
This type of restriction is commonly used under flexible loss and is similar to Assumption A.1 of \cite{lu2015jackknife} and C.1 of \cite{bai2022optimal}. 
These moment conditions imply that
$\operatorname{E}\left\|\mathbf{x}_{i(m)}\right\|^{ s} \leq c_{\mathbf{x}} k_{m}^{s/2}, m=1, 2,\ldots M, s=1,\ldots,16 $ and are crucial in subsequent proofs, such as in the application of Bernstein and Markov inequalities.
\begin{assumption}
\label{assumption2}
\begin{enumerate}[(i)]
	\item 
	$  
	\max _{1 \leq m \leq M}\left\|\hat{\boldsymbol{\theta}}_{(m)}-\boldsymbol{\theta}_{(m)}^{*}\right\|=O_p(c_n);
	$
	\item
	$  \max_{1 \leq j \leq J}
	\max _{1 \leq m \leq M}\left\|\tilde{\boldsymbol{\theta}}_{(m)}^{[-j]}-\boldsymbol{\theta}_{(m)}^{*}\right\|=O_p(c_n).
	$
\end{enumerate}
\end{assumption}

\cref{assumption2} is a high-level condition, which ensures that $\hat{\boldsymbol{\theta}}_{(m)}$ and $\tilde{\boldsymbol{\theta}}_{(m)}^{[-j]}$ uniformly converge to $\boldsymbol{\theta}_{(m)} ^{*}$ at a rate of $c_n$. Note that  $c_n$ is determined by the sample size $n$ and largest dimension of the models $\bar{k}$. Additionally, from \cite{lu2015jackknife} and \cite{tu2020jackknife}, we have $c_n=n^{-1/2}\bar{k}^{1/2}(\log n )^{1/2}$ for diverging $M$ and $c_n=n^{-1/2}\bar{k}^{1/2}$ for finite $M$ under some regularity conditions. This $c_n$ may be improved, but we leave this for future research.

\begin{assumption}
\label{assumption3}
\begin{enumerate}[(i)]
    \item  $\xi_n^{-1} c_n\bar{k}^{p-1/2} =o(1) $,
\item  $n^{-1}\bar{k}^{2p}M \left[\log(\bar{k}^p\log n)-\log \xi_n\right]  =o(1)$,
\item  $\xi_n^{-4} n^{-3}  M^5 \bar{k}^{4p} \left[\log(\bar{k}^p\log n)-\log \xi_n\right]^4  =o(1)$.
\end{enumerate}
\end{assumption}

\cref{assumption3} imposes restrictions on the largest dimension of the models $\bar{k}$, the number of candidate models $M$ and $\xi_n$. Furthermore, it requires that all candidate models be misspecified because we need $\xi_n\neq 0$  for all $n$ (otherwise, $\xi_n\leq \operatorname{E}\left[\rho_{\tau,p}\left(y_{n+1}-\mu_{n+1}\right)\right]-
\operatorname{E}\left[\rho_{\tau,p}\left(\varepsilon_{n+1}\right)\right]=0 $). Nevertheless, we allow $\xi_n$ to go to zero at a much slower rate than $n$.

The following theorem states that our JCVMA-selected weight vector $\hat{\mathbf{w}}$ is asymptotically optimal in terms of minimizing the EFPE. That is, among all feasible weight vectors, $\hat{\mathbf{w}}$ makes $\operatorname{EFPE}_{n}(\mathbf{w})$ as small as possible when $n\to \infty$.
\begin{theorem}
\label{thm:opt_EFPE}
If  Assumptions \ref{assumption1}, \ref{assumption2}, and \ref{assumption3}  hold,
then $\hat{\mathbf{w}}$ is asymptotically optimal, in the sense that
\begin{equation} 
\label{eq:opt_EFPE}
	\frac{\operatorname{EFPE}(\hat{\mathbf{w}})}{\inf _{\mathbf{w} \in \mathcal{W}} \operatorname{EFPE}(\mathbf{w})}\stackrel{p}{\longrightarrow} 1.
\end{equation}
\end{theorem}

Note that \cite{lu2015jackknife}, \cite{wang2023jackknife}, and \cite{tu2020jackknife} established asymptotic optimality by minimizing the FPE instead of the EFPE. Our asymptotic optimality in \cref{thm:opt_EFPE} improves upon their results by using a more appropriate prediction error measure.
Moreover, we obtained a similar result in the following theorem, which shows the asymptotic optimality of $\hat{\mathbf{w}}$ in terms of minimizing the FPE.

We make the following assumption as an alternative to \cref{assumption3}, which is used in the following theorem.
\begin{assumption}
\label{assumption4}
\begin{enumerate}[(i)]
    \item $ c_n\bar{k}^{p-1/2}=o(1)$,
\item  $n^{-1}\bar{k}^{2p}M\log(\bar{k}^p\log n)=o(1)$,
\item  $ n^{-3}  M^5 \bar{k}^{4p} \left[\log(\bar{k}^p\log n)\right]^4  =o(1)$.
\end{enumerate}
\end{assumption}
\cref{assumption4} imposes restrictions on the largest dimension of the models $\bar{k}$ and the number of candidate models $M$. As $\operatorname{FPE}(\mathbf{w}) \geq \operatorname{E}\left[\rho_{\tau,p}\left(\varepsilon_{n+1}\right)\right] > 0$, we need not
consider the infimum of FPE. Therefore, it does not require the misspecification of candidate models.

Then, we have the following theorem, which shows that our JCVMA-selected weight vector $\hat{\mathbf{w}}$ is asymptotically optimal in terms of minimizing the FPE. 
\begin{theorem}
\label{thm:opt_FPE}
If Assumptions \ref{assumption1}, \ref{assumption2}, and \ref{assumption4} hold,
then $\hat{\mathbf{w}}$ is asymptotically optimal in the sense that
\begin{equation} 
\label{eq:opt_FPE}
\frac{\operatorname{FPE}(\hat{\mathbf{w}})}{\inf _{\mathbf{w} \in \mathcal{W}} \operatorname{FPE}(\mathbf{w})}\stackrel{p}{\longrightarrow} 1.
\end{equation}
\end{theorem}
Because $\operatorname{FPE}_{n}(\mathbf{w})-\operatorname{EFPE}_{n}(\mathbf{w})=\operatorname{E}\left[\rho_{\tau,p}\left(\varepsilon_{n+1}\right)\right]$ is a constant, we can obtain \eqref{eq:opt_FPE} from \eqref{eq:opt_EFPE}. Specifically, if \eqref{eq:opt_EFPE} holds, we have the following: 
\begin{align}
&\frac{\operatorname{FPE}(\hat{\mathbf{w}})-\inf _{\mathbf{w} \in \mathcal{W}} \operatorname{FPE}(\mathbf{w})}{\inf _{\mathbf{w} \in \mathcal{W}} \operatorname{FPE}(\mathbf{w})} \notag \\ 
={}& \frac{\operatorname{EFPE}(\hat{\mathbf{w}})-\inf _{\mathbf{w} \in \mathcal{W}} \operatorname{EFPE}(\mathbf{w})}{\inf _{\mathbf{w} \in \mathcal{W}} \operatorname{EFPE}(\mathbf{w})} 
\frac{\inf _{\mathbf{w} \in \mathcal{W}} \operatorname{EFPE}(\mathbf{w})}{\inf _{\mathbf{w} \in \mathcal{W}} \operatorname{FPE}(\mathbf{w})}\notag \\
= {}& o_p(1).
\end{align}
Therefore, the asymptotic optimality of \cref{thm:opt_FPE} is weaker than that of \cref{thm:opt_EFPE}. However, compared with \cref{thm:opt_EFPE}, \cref{thm:opt_FPE} does not require all candidate models to be misspecified.
Note that \eqref{eq:opt_FPE} makes sense only if neither of the aforementioned situations (i.e., [i] or [ii])  is satisfied. The proof of \cref{thm:opt_FPE}
 is similar to that of \cref{thm:opt_EFPE} and, thus, omitted.

\subsection{Estimation Consistency}

Here, we consider the case where the set of candidate models includes at least one correct model but not necessarily the true model. Let $\mathcal{D} $ be the subset of $\{ 1, \ldots, M \}$ that comprises the correct models. Further, assume that $\mathcal{D}$ is not empty.

Next, assume that $\mu_i$ is given by $\mu_i =\mathbf{x}_{i,0}^{\prime} \boldsymbol{\Theta}_0 $, where $\mathbf{x}_{i,0}=(x_{i1},\ldots,x_{i \bar{k}})^{\prime} $ and 
$\boldsymbol{\Theta}_{0} =\left(\theta_{1},  \ldots,  \theta_{\bar{k}}\right)^{\prime}$. 
Let $\Pi_{(m)}$ be the selection matrix of the $m$th model, such that $\Pi_{(m)}=(\mathbf{I}_{k_{m}}, \mathbf{0}_{k_{m} \times(\bar{k}-k_{m})})$ or a column permutation thereof, where $\mathbf{I}_{k_{m}}$ denotes a $k_m \times k_m$ identity matrix and $\mathbf{0}_{k_{m} \times(\bar{k}-k_{m})}$ denotes a $k_{m} \times\left(\bar{k}-k_{m}\right)$ matrix of zeros. Under the $m$th candidate model, we denote $\hat{\boldsymbol{\Theta}}_{(m)}= \Pi_{(m)}^{\prime} \hat{\boldsymbol{\theta}}_{(m)} $ as the  estimator of the coefficient $\boldsymbol{\Theta}_{0} $, which is a $\bar{k}\times 1 $ vector. Therefore, the model averaging estimator of $\boldsymbol{\Theta}_{0} $ is given by
$$
\hat{\boldsymbol{\Theta}}(\mathbf{w})=\sum_{m=1}^{M}w_m \hat{\boldsymbol{\Theta}}_{(m)},
$$
and
the model averaging estimator of $\mu_i$ becomes the following:
$
\hat{\mu}_{i}(\mathbf{w})=\sum_{m=1}^{M} w_{m} \mathbf{x}_{i,0}^{\prime} \hat{\boldsymbol{\Theta}}_{(m)}
=\mathbf{x}_{i,0}^{\prime}\hat{\boldsymbol{\Theta}}(\mathbf{w}).
$
Similarly, denote $
\tilde{\boldsymbol{\Theta}}_{(m)}^{[-j]}=\Pi_{(m)}^{\prime} \tilde{\Theta}_{(m)}^{[-j]}$ and
$\tilde{\boldsymbol{\Theta}}^{[-j]}(\mathbf{w})=\sum_{m=1}^{M} w_{m} \tilde{\boldsymbol{\Theta}}_{(m)}^{[-j]}.
$ 
Then,
$\tilde{\mu}_{(j-1)Q+q}^{[-j]}(\mathbf{w}) = \sum_{m=1}^{M} w_{m} \mathbf{x}_{(j-1)Q+q,0}^{ \prime} \tilde{\boldsymbol{\Theta}}_{(m)}^{[-j]}
=\mathbf{x}_{(j-1)Q+q,0}^{ \prime}\tilde{\boldsymbol{\Theta}}^{[-j]}(\mathbf{w}) .
$

To ensure the consistency of the coefficient model averaging estimator $\hat{\boldsymbol{\Theta}}(\mathbf{w})$, we impose the following assumption.
\begin{assumption}
\label{assumption5}
For any $m$ in $\mathcal{D}$, we have:
\begin{enumerate}[(i)]
\item 
$  
	\left\|\hat{\boldsymbol{\Theta}}_{(m)}-\boldsymbol{\Theta}_{0}\right\|=O_p(n^{-1/2}\bar{k}^{1/2});
	$
\item $  
	\left\|\tilde{\boldsymbol{\Theta}}^{[-j]}_{(m)}-\boldsymbol{\Theta}_{0}\right\|=O_p(n^{-1/2}\bar{k}^{1/2})
	$ uniformly for $j=1,\ldots,J.$
\end{enumerate}
\end{assumption}
\cref{assumption5} states the convergence rate of the estimator of the coefficient $\boldsymbol{\Theta}_{0} $ under the correct model, which is derived under some regularity conditions (e.g., \citealt{lu2015jackknife,bai2022optimal}).

\begin{assumption}
\label{assumption6}
There exist constants $ 0< \kappa_1\leq \kappa_2<\infty$ such that $$ \kappa_1\leq\lambda_{\min}\left(n^{-1}\sum_{i=1}^{n} 
\mathbf{x}_{i,0}\mathbf{x}_{i,0}^{\prime}\right)\leq 
\lambda_{\max}\left(n^{-1}\sum_{i=1}^{n} 
\mathbf{x}_{i,0}\mathbf{x}_{i,0}^{\prime}\right) \leq \kappa_2 ,$$ and
$$\kappa_1\leq \lambda_{\min}\left(Q^{-1}\sum_{q=1}^{Q}. 
\mathbf{x}_{(j-1)Q+q,0}\mathbf{x}_{(j-1)Q+q,0}^{\prime}\right) \leq 
\lambda_{\max}\left(Q^{-1}\sum_{q=1}^{Q} ,  
\mathbf{x}_{(j-1)Q+q,0}\mathbf{x}_{(j-1)Q+q,0}^{\prime}\right) \leq \kappa_2  $$ a.s. uniformly for $j=1,\ldots,J$.
\end{assumption}

\cref{assumption6} is used in \cite{zhang2020parsimonious}, who assumed the predictors have reasonably good behavior.

\begin{assumption}
\label{assumption7}
\begin{enumerate}[(i)]
    \item There exists a constant $\varrho>0$, such that $|\mu_i-\mu_i^*(\mathbf{w})|< \varrho $ a.s. uniformly for $i =1, \ldots, n$ and $\mathbf{w}\in \mathcal{W} $.
\item  There exist constants $0<\delta_1 \leq \delta_2 < \infty $, such that $\delta_1 \leq f\left(s \mid \mathbf{x}_{i}\right) \leq \delta_2 $ a.s. uniformly for $|s|\leq \varrho$ and $i =1, \ldots, n$.
\end{enumerate}
\end{assumption}
\cref{assumption7} is used in \cite{xumodel}, which is mild and excludes some pathological cases in which $\mu_i-\mu_i^*(\mathbf{w})$ explodes.

\begin{theorem}
\label{thm:estconsistency}
Suppose that $\mathcal{D}$ is not empty.
\begin{enumerate}[(i)]
\item For $p=1$, if Assumptions \ref{assumption1}, \ref{assumption2}, \ref{assumption5}, \ref{assumption6}, and \ref{assumption7} hold,
then the model averaging estimator $\hat{\boldsymbol{\Theta}}(\hat{\mathbf{w}})$ satisfies
$$
\left\|\hat{\boldsymbol{\Theta}}(\hat{\mathbf{w}})-\boldsymbol{\Theta}_{0}\right\|=O_p(c_n^{1/2}\bar{k}^{1/4}+n^{-1/4}M^{1/4}).
$$
\item For  $p=2$, if Assumptions \ref{assumption1}, \ref{assumption2}, \ref{assumption5}, and \ref{assumption6} hold,
then the model averaging estimator $\hat{\boldsymbol{\Theta}}(\hat{\mathbf{w}})$ satisfies

$$
\left\|\hat{\boldsymbol{\Theta}}(\hat{\mathbf{w}})-\boldsymbol{\Theta}_{0}\right\|=O_p(c_n).
$$
\end{enumerate}
\end{theorem}

\cref{thm:estconsistency} establishes a convergence rate for the model averaging estimator $\hat{\boldsymbol{\Theta}}(\hat{\mathbf{w}})$, which is determined by $c_n$, $M$, and $\bar{k}$. If $c_n^{1/2}\bar{k}^{1/4}$ and $n^{-1/4}M^{1/4}$ approach 0 as $n$ increases, we can obtain the consistency of  $\hat{\boldsymbol{\Theta}}(\hat{\mathbf{w}})$.

\subsection{Weight Convergence}
The third justification is that the proposed averaging prediction asymptotically assigns all weights to the correct models included in the model set.  Here, we consider a case where $\mathcal{D}$ is not empty. Let $\hat{\tau}=\sum_{m \in \mathcal{D}} \hat{w}_{m}$ be the sum of the JCVMA-selected weights assigned to the correct models.
We aim to demonstrate that $\hat{\tau} \rightarrow 1$ in probability under certain regularity conditions.

Let $\mathcal{W}_{F}=\left\{\mathbf{w} \in \mathcal{W}: \sum_{m \notin \mathcal{D}} w_{m}=1\right\}$ be the subset of $\mathcal{W}$ that assigns all weights to the misspecified model. The following assumption is imposed for the case in which some models are specified correctly.

\begin{assumption}
\label{assumption8}
There exists a constant $c_0$, such that
$\inf _{\mathbf{w} \in \mathcal{W}_{F}} \operatorname{E}\left[\left(
\mu^*_{n+1}(\mathbf{w})-\mu_{n+1}\right)^2\right]\geq c_0 >0. $
\end{assumption}
\cref{assumption8} is similar to Assumption 5 in \cite{yu2022unified}. It requires that some candidate models are misspeciﬁed and the corresponding misspeciﬁcation error does not vanish as $n \rightarrow \infty$.

\begin{theorem}
\label{thm:weightconvergence}
Suppose that $\mathcal{D}$ is not empty.
\begin{enumerate}[(i)]
\item For $p=1$, if Assumptions \ref{assumption1}, \ref{assumption2}, \ref{assumption4}, \ref{assumption7}, and \ref{assumption8} hold,
then we have $\hat{\tau} \stackrel{p}{\longrightarrow} 1$.
\item For  $p=2$, if Assumptions \ref{assumption1}, \ref{assumption2}, \ref{assumption4}, and \ref{assumption8} hold,
then we have $\hat{\tau} \stackrel{p}{\longrightarrow} 1$.
\end{enumerate}
\end{theorem}
\cref{thm:weightconvergence} shows that the proposed averaging method asymptotically assigns all weights to the correct models included in the model set. This result corresponds to the consistency property in model selection.

\begin{remark}
\cref{thm:estconsistency} and \cref{thm:weightconvergence} are different. The former concerns parameter estimation, while the latter concerns averaging weights. Nevertheless, both are important.
\end{remark}

\subsection{Examples}
\label{sec:examples}
\subsubsection{Quantile Regression Model Averaging.}	
When $p=1$, $\rho_{\tau,1}(\lambda)=[\tau-\mathbf{1}\{\lambda \leq 0\}]\lambda$, which is an asymmetric linear loss function, and our model averaging method becomes QR model averaging.
The following Corollaries \ref{cor:opt_EFPE_qr} and \ref{cor:opt_FPE_qr} state that our JCVMA-selected weight vector $\hat{\mathbf{w}}$ is asymptotically optimal for minimizing EFPE and FPE, respectively.
\begin{corollary}
\label{cor:opt_EFPE_qr}
For $p=1$, if  Assumptions \ref{assumption1}, \ref{assumption2}, and \ref{assumption3}  hold,
then $\hat{\mathbf{w}}$ is asymptotically optimal, as
$$	\frac{\operatorname{EFPE}(\hat{\mathbf{w}})}{\inf _{\mathbf{w} \in \mathcal{W}} \operatorname{EFPE}(\mathbf{w})}\stackrel{p}{\longrightarrow} 1.
$$
\end{corollary}

\begin{corollary}
\label{cor:opt_FPE_qr}
For $p=1$, if  Assumptions \ref{assumption1}, \ref{assumption2}, and \ref{assumption5} hold,
then $\hat{\mathbf{w}}$ is asymptotically optimal, as
$$
	\frac{\operatorname{FPE}(\hat{\mathbf{w}})}{\inf _{\mathbf{w} \in \mathcal{W}} \operatorname{FPE}(\mathbf{w})}\stackrel{p}{\longrightarrow} 1.
$$
\end{corollary}
Note that almost all studies on QR model averaging established asymptotic optimality by minimizing the FPE, such as \cite{lu2015jackknife} and \cite{wang2023jackknife}. Meanwhile, our asymptotic optimality in \cref{cor:opt_EFPE_qr} improves these results.
Moreover, to the best of our knowledge, no study has shown the estimation consistency and weight convergence of the QR model averaging estimator. We prove the convergence rate of $\hat{\boldsymbol{\Theta}}(\hat{\mathbf{w}})$ and the weight convergence of our QR model averaging estimator, which 
are shown in the Corollaries \ref{cor:consistency_QR} and \ref{cor:weightconvergence_QR}, respectively.

\begin{corollary}
\label{cor:consistency_QR}
We assume that $\mathcal{D}$ is not empty.
For $p=1$,	if Assumptions \ref{assumption1}, \ref{assumption2}, \ref{assumption5}, \ref{assumption6}, and \ref{assumption7} hold,
then the QR model averaging estimator $\hat{\boldsymbol{\Theta}}(\hat{\mathbf{w}})$ satisfies:
$$
\left\|\hat{\boldsymbol{\Theta}}(\hat{\mathbf{w}})-\boldsymbol{\Theta}_{0}\right\|=O_p(c_n^{1/2}\bar{k}^{1/4}+n^{-1/4}M^{1/4}).
$$
\end{corollary}

\begin{corollary}
\label{cor:weightconvergence_QR}
We assume that $\mathcal{D}$ is not empty.
	For $p=1$, if Assumptions \ref{assumption1}, \ref{assumption2}, \ref{assumption4}, \ref{assumption7}, and \ref{assumption8} hold,
then we get $\hat{\tau} \stackrel{p}{\longrightarrow} 1$.
\end{corollary}
Note that if $c_n^{1/2}\bar{k}^{1/4}$ and $n^{-1/4}M^{1/4}$ approach 0 as $n$ increases, we can obtain the consistency of  $\hat{\boldsymbol{\Theta}}(\hat{\mathbf{w}})$ from \cref{cor:consistency_QR}.

\subsubsection{Expectile Regression Model Averaging.}

When $p=2$, $\rho_{\tau,2}(\lambda)=|\tau-\mathbf{1}\{\lambda \leq 0\}|\lambda^2$, which is an asymmetric quadratic loss function, while our model averaging method becomes ER model averaging.
The Corollaries \ref{cor:opt_EFPE_er} and \ref{cor:opt_FPE_er} state that our JCVMA-selected weight vector $\hat{\mathbf{w}}$ is asymptotically optimal for minimizing the EFPE and FPE, respectively.
\begin{corollary}
\label{cor:opt_EFPE_er}
For $p=2$, if  Assumptions \ref{assumption1}, \ref{assumption2}, and \ref{assumption3}  hold,
then $\hat{\mathbf{w}}$ is asymptotically optimal, in the sense that
$$	\frac{\operatorname{EFPE}(\hat{\mathbf{w}})}{\inf _{\mathbf{w} \in \mathcal{W}} \operatorname{EFPE}(\mathbf{w})}\stackrel{p}{\longrightarrow} 1.
$$
\end{corollary}

\begin{corollary}
\label{cor:opt_FPE_er}
For $p=2$, if  Assumptions \ref{assumption1}, \ref{assumption2}, and \ref{assumption5} hold,
then $\hat{\mathbf{w}}$ is asymptotically optimal in the sense that
$$
	\frac{\operatorname{FPE}(\hat{\mathbf{w}})}{\inf _{\mathbf{w} \in \mathcal{W}} \operatorname{FPE}(\mathbf{w})}\stackrel{p}{\longrightarrow} 1.
$$
\end{corollary}

Note that research on ER model averaging has typically established asymptotic optimality by minimizing the FPE, such as the study of \cite{tu2020jackknife}. Meanwhile, our asymptotic optimality in \cref{cor:opt_EFPE_er} improves upon these results.
Moreover, the Corollaries \ref{cor:consistency_ER} and  \ref{cor:weightconvergence_ER} show 
the convergence rate of $\hat{\boldsymbol{\Theta}}(\hat{\mathbf{w}})$ and
the weight convergence of our ER model averaging estimator, respectively. 

\begin{corollary}
\label{cor:consistency_ER}
Suppose that $\mathcal{D}$ is not empty.
For  $p=2$, if Assumptions \ref{assumption1}, \ref{assumption2}, \ref{assumption5} and \ref{assumption6} hold,
then the model averaging estimator $\hat{\boldsymbol{\Theta}}(\hat{\mathbf{w}})$ satisfies

$$
\left\|\hat{\boldsymbol{\Theta}}(\hat{\mathbf{w}})-\boldsymbol{\Theta}_{0}\right\|=O_p(c_n).
$$
\end{corollary}

\begin{corollary}
\label{cor:weightconvergence_ER}
Suppose that $\mathcal{D}$ is not empty.
For  $p=2$, if Assumptions \ref{assumption1}, \ref{assumption2}, \ref{assumption4} and \ref{assumption8} hold,
then we have $\hat{\tau} \stackrel{p}{\longrightarrow} 1$.
\end{corollary}
Note that if $c_n$ approaches 0 as $n$ increases, we can obtain the consistency of  $\hat{\boldsymbol{\Theta}}(\hat{\mathbf{w}})$ from 
\cref{cor:consistency_ER}.

\subsection{Simulation Evidence}	
\label{sec:simulation}
In this subsection, we conduct Monte Carlo experiments to evaluate the finite sample performance of the proposed JCVMA method. We consider two simulation designs. In the first design, all candidate models are misspecified. In the second design, at least one correct model is included in the set of candidate models.

\subsubsection{Simulation Design I.}
\label{subsec:sd1}
Simulation Design I examines how the proposed model averaging method performs when the data are generated from a model that is not included in the candidate set. 
We consider the following three DGPs. The first DGP is linear and considers nested models with divergent dimensions, the second one is linear and considers non-nested models with fixed dimensions, and the third one is nonlinear and considers nested models with divergent dimensions.

The first DGP is obtained from the research of \cite{lu2015jackknife}:

\centerline{ $\text{DGP 1: }  y_{i}=\theta \sum_{j=1}^{1000} j^{-1} x_{i j}+\varepsilon_{i},$ \quad (linear, nested and divergent $M$)}

\noindent where $x_{i 1}=1$ and $x_{i j}, j=2,3, \ldots,1000$ are IID $N(0,1)$ and mutually independent of each other. We only consider heteroskedasticity, where $\varepsilon_{i}=\sum_{j=2}^{6} x_{i j}^{2}\epsilon_{i}$, while $\epsilon_i$ is $N(0,1)$ and independent of $x_{i j}, j=2,3, \ldots,1000$. 
This DGP is an approximation of \eqref{eq:DGP}, which is a high-dimensional regression model. Unlike \cite{lu2015jackknife}, we set the number of models $M=\left\lfloor 5 n^{1 / 5}\right\rfloor$, which increases with the sample size $n$. We consider $M$ nested models by specifying $\mathbf{x}_{i(1)}=\left(x_{i 1}\right)', \mathbf{x}_{i(2)}=\left(x_{i 1}, x_{i 2}\right)'$, etc.

The second DGP is as follows:

\centerline{\noindent $\text{DGP 2: } y_{i}=\theta \sum_{j=1}^{30}\beta_j x_{i j}+\varepsilon_{i},$ \quad (linear, non-nested and fixed $M$)}

\noindent where $x_{i 1}=1$, $x_{i j}, j=2,3, \ldots,30$ are IID $N(0,1)$ and mutually independent of each other, and
$(\beta_1,\beta_2,\ldots,\beta_{30})=(1,1,1,0,0,1,2,3,1,1,\ldots,1) $.
We set $\varepsilon_{i}=\sum_{j=2}^{8}  x_{i j}^2 \epsilon_{i}$, where $\epsilon_i$ is $N(0,1)$ and independent of $x_{i j}, j=2,3, \ldots,30$. 
Compared with DGP 1, this DGP reduces the dimension of the regression model and the coefficients do not decay. We assume that the first eight covariates are observable, where the first three variables must be included in the candidate models and the last five uncertain variables may not be included in the candidate models. Consider non-nested candidate models; that is, $M=2^5=32$, where $M$ is fixed.

The third DGP is obtained from the study of \cite{lu2015jackknife}:

\centerline{ $\text{DGP 3: } y_{i}=\theta\left[x_{i 1}+\sum_{j=2}^{25} j^{-1} \Phi\left(x_{i j}\right)\right]+\varepsilon_{i},$ \quad (nonlinear, nested and divergent $M$)}

\noindent 
where $x_{i 1}=1$, $x_{i j}, j=2,3, \ldots,25$ are IID $N(0,1)$ and mutually independent of each other, and $\Phi(\cdot)$ is the standard normal CDF. We set $\varepsilon_{i}=\left(0.01+\sum_{j=2}^{11} x_{i j}^{2}\right) \epsilon_{i}$, where $\epsilon_i$ is $N(0,1)$ and independent of $x_{i j}, j=2,3, \ldots,25$. Unlike DGPs 1 and 2, DGP 3 is nonlinear.  Unlike \cite{lu2015jackknife}, we consider $M=\left\lfloor 5 n^{1 / 5}\right\rfloor$ nested models by specifying $\mathbf{x}_{i(1)}=\left(x_{i 1}\right)', \mathbf{x}_{i(2)}=\left(x_{i 1}, x_{i 2}\right)'$, etc.

We use the population $R^{2}=\left[ \operatorname{var}\left(y_{i}\right)-\operatorname{var}\left(\varepsilon_{i}\right)\right] / \operatorname{var}\left(y_{i}\right)$ to control $\theta$, such that $R^2=0.1,0.2,\ldots,0.9$.
For $n = 50, 100, 200,400, p=1,2$, and $\tau=0.5,0.05$, we consider the following model averaging methods:
\begin{enumerate}[(1)]
\item JCVMA proposed in Section \ref{sec:MA} with $J=5$ (JCVMA5);
\item JCVMA with $J=10$ (JCVMA10);
\item Smoothed Akaike information criterion model averaging (SAIC);
\item Smoothed Bayesian information criterion  model averaging (SBIC); 
and
\item Equal weight averaging (EWA).
\end{enumerate}
\begin{remark}
We also consider the corresponding model selection methods, namely, the Akaike information criterion (AIC), the Bayesian information criterion (BIC), 5-fold CV, and 10-fold CV. As none of them perform better than the corresponding model averaging methods, the results are not shown.
\end{remark}

Following \cite{demetrescu2019predictive}, for the $m$th model with $k_m$ regressors, AIC and BIC are defined as follows:
$$
\begin{aligned}
&A I C_{m}=\frac{2}{p} n \log \left[\frac{1}{n} \sum_{i=1}^{n} \rho_{\tau,p}\left(y_{i}-\hat{\Theta}_{(m)}^{\prime} \mathbf{x}_{i(m)}\right)\right]+2 k_{m}, ~\text{and} \\
&B I C_{m}=\frac{2}{p} n \log \left[\frac{1}{n} \sum_{i=1}^{n} \rho_{\tau,p}\left(y_{i}-\hat{\Theta}_{(m)}^{\prime} \mathbf{x}_{i(m)}\right)\right]+k_{m} \ln (n) .
\end{aligned}
$$
Then, following \cite{buckland1997model},   SAIC and SBIC weights for the $m$th model are obtained as follows:
$$
\hat{w}_{m}^{A I C}=\frac{\exp \left(-A I C_{m} / 2\right)}{\sum_{m=1}^{M} \exp \left(-A I C_{m} / 2\right)} ~\text{and}~ \hat{w}_{m}^{B I C}=\frac{\exp \left(-B I C_{m} / 2\right)}{\sum_{m=1}^{M} \exp \left(-B I C_{m} / 2\right)}.
$$

For the $r$th  replication, we generate $\left\{(y_{n+s}^{(r)}, \mathbf{x}_{n+s}^{(r)})\right\}_{s=1}^{100}$ from each DGP as out-of-sample observations. The excess final prediction error is calculated as follows: 
$$
\operatorname{EFPE}(r)=\frac{1}{100} \sum_{s=1}^{100} \rho_{\tau,p}\left(y_{n+s}^{(r)}-\hat{\mu}_{n+s}^{(r)} \right)- \operatorname{E}\left[\rho_{\tau,p}\left(\varepsilon_{n+1}^{(r)}-\arg \min _{q \in \mathbb{R}} \operatorname{E}\left[ \rho_{\tau, p}\left(\varepsilon_{n+1}^{(r)}-q
 \right)\mid \mathbf{x}_{n+1}^{(r)}\right]\right)\right],
$$
where $\hat{\mu}_{n+s}^{(r)}$ is determined using each method. Then, we average the excess out-of-sample prediction error over $R = 500$ replications:
$
\mathrm{EFPE}=\frac{1}{R} \sum_{r=1}^{R} \mathrm{EFPE}(r).
$
The smaller the EFPE, the better the method in terms of excess out-of-sample prediction error. Following \cite{hansen2007least}, we normalize the EFPE by dividing by the EFPE of the infeasible optimal single model.

\begin{figure}[htbp]
\centering
\includegraphics[scale=0.52]{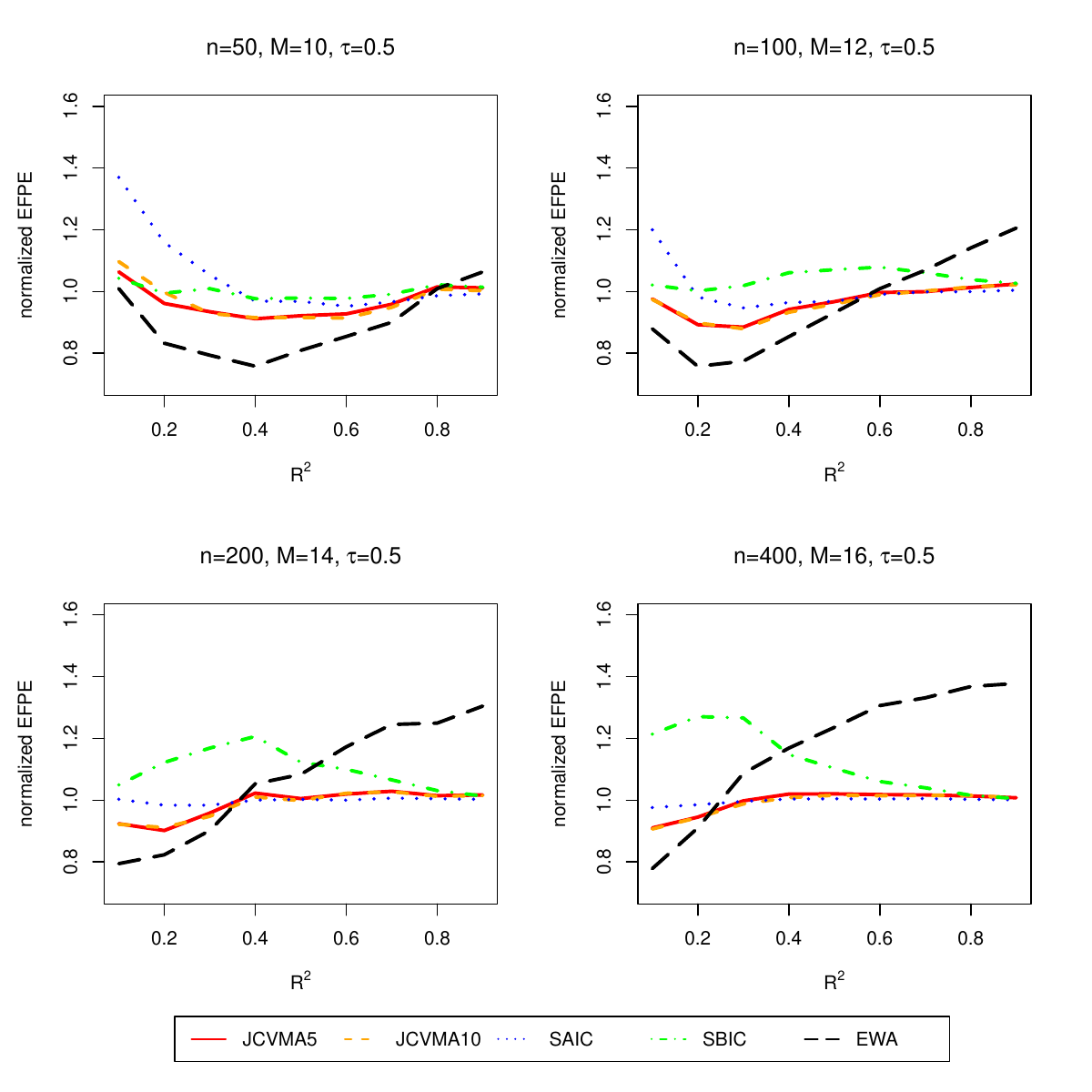}
\caption{Normalized EFPE: DGP 1, $p=1,\tau=0.5$}
\label{fig:DGP1_QR_0.5}
\end{figure}

\begin{figure}[htbp]
\centering
\includegraphics[scale=0.52]{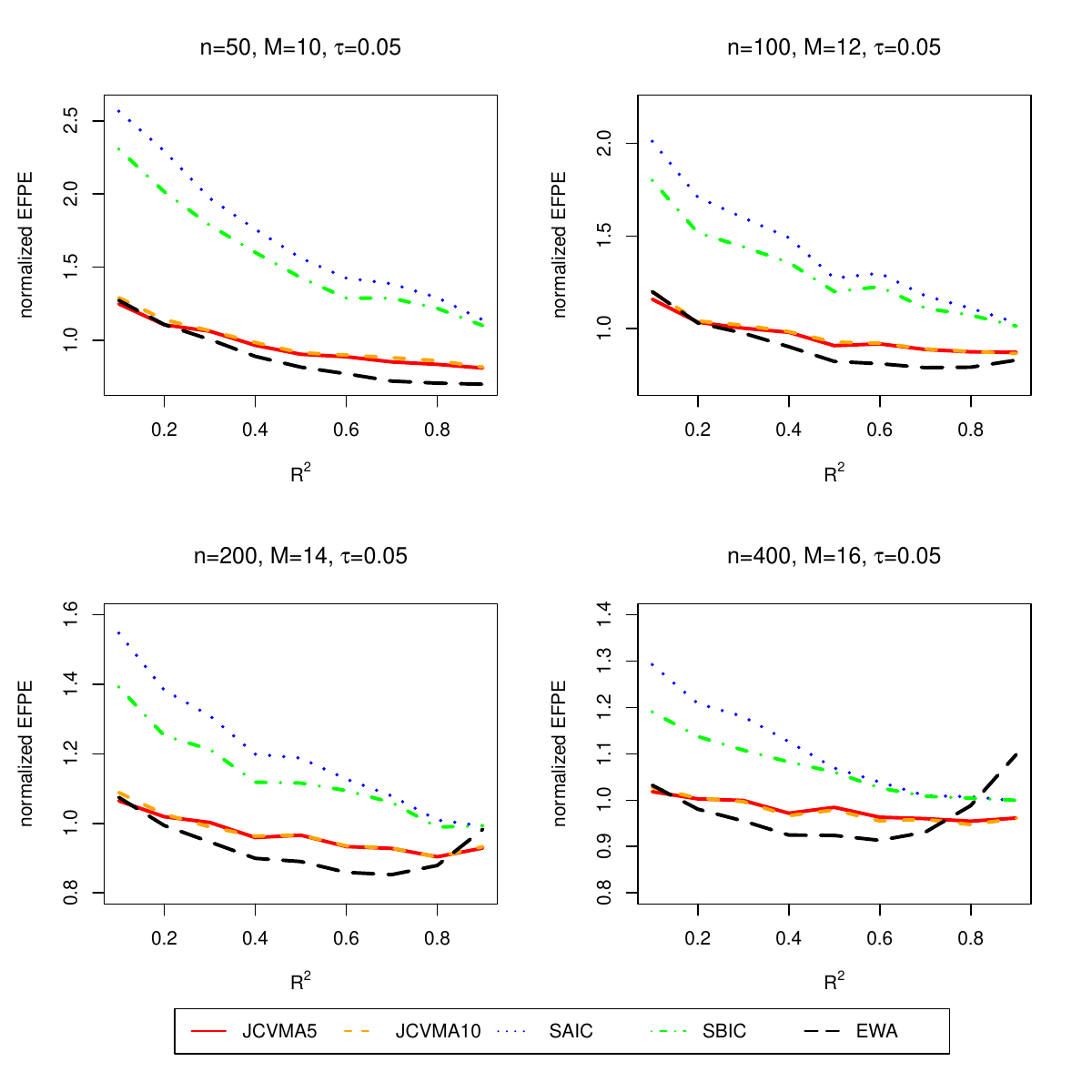}
\caption{Normalized EFPE: DGP 1, $p=1,\tau=0.05$}
\label{fig:DGP1_QR_0.05}
\end{figure}

\begin{figure}[htbp]
\centering
\includegraphics[scale=0.52]{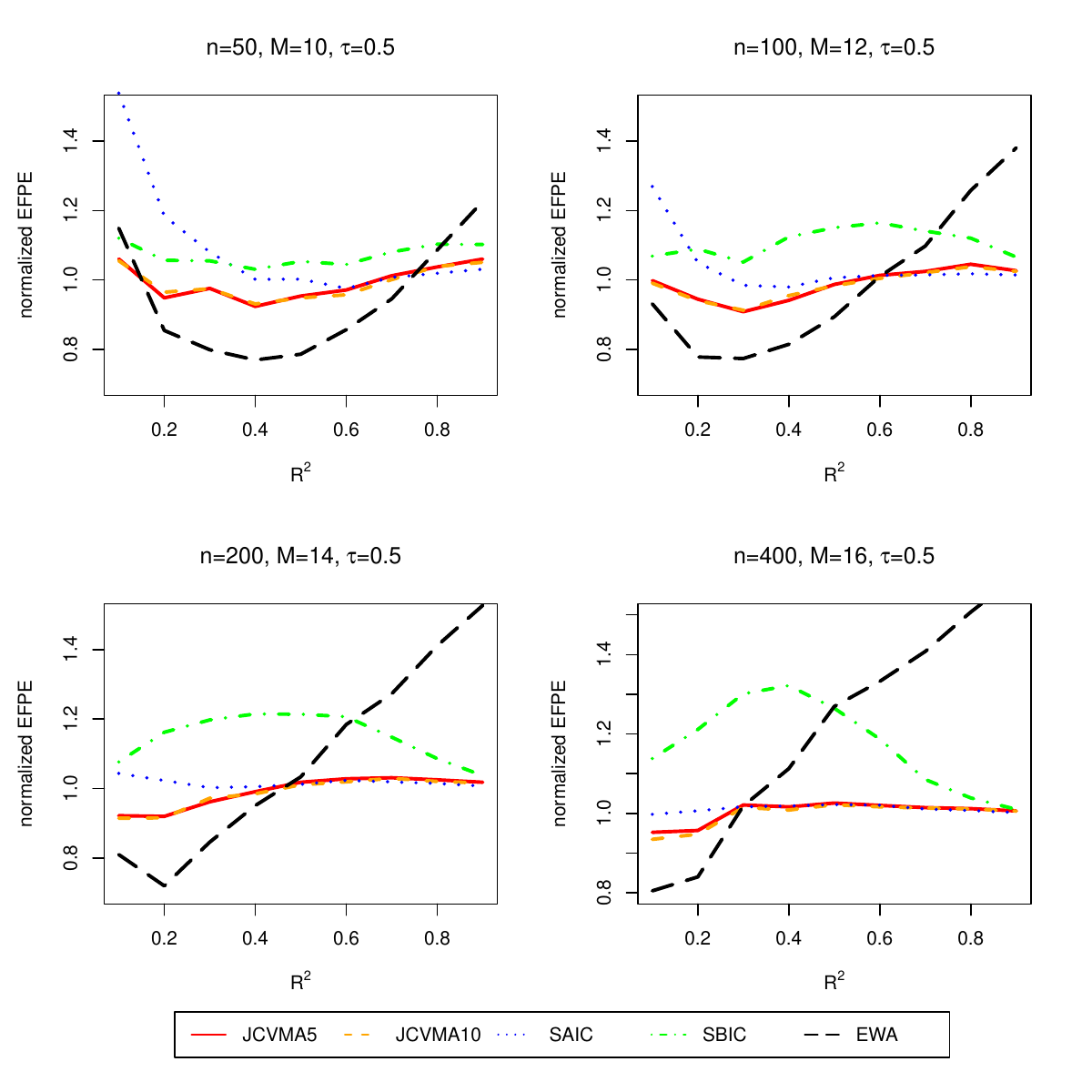}
\caption{Normalized EFPE: DGP 1, $p=2,\tau=0.5$}
\label{fig:DGP1_ER_0.5}
\end{figure}

\begin{figure}[htbp]
\centering
\includegraphics[scale=0.52]{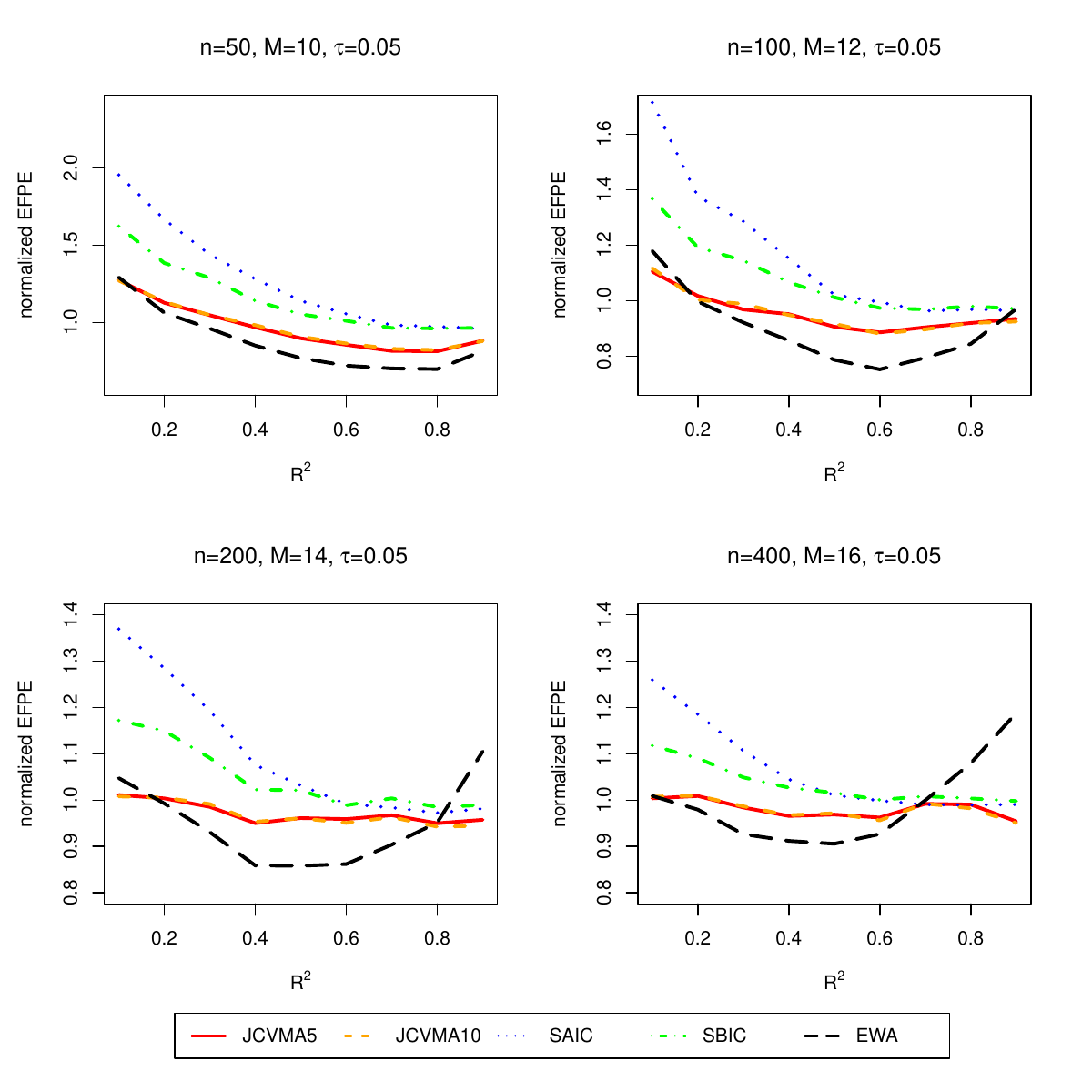}
\caption{Normalized EFPE: DGP 1, $p=2,\tau=0.05$}
\label{fig:DGP1_ER_0.05}
\end{figure}

Figures \ref{fig:DGP1_QR_0.5}, \ref{fig:DGP1_QR_0.05}, 
\ref{fig:DGP1_ER_0.5}, and \ref{fig:DGP1_ER_0.05}
visualize the relationship between the normalized FPE and $R^2$ under different conﬁgurations of $p,\tau,$ and $n$.
Overall, JCVMA outperforms  SAIC and SBIC, and is much more stable than EWA. JCVMA5 is almost indistinguishable from JCVMA10.
Specifically, when $p=1,\tau=0.5$, while $n$ is very small, JCVMA is better than  SAIC and SBIC, and inferior to EWA. However, EWA deteriorates sharply with increasing $n$, especially for large $R^2$, whereas JCVMA shows stable performance. When $p=1,\tau=0.05$, JCVMA is obviously better than SAIC and SBIC, while EWA deteriorates with an increase in $n$ for large $R^2$. Similar results are obtained for $p=2$.
Together, this shows that when $n$ is very small, instead of using complex methods to choose weights, using the simplest equal-weight averaging to make predictions can be a better choice.
When $n$ is large, the advantage of JCVMA, which has asymptotic optimality, is evident.

The simulation results of DGP 2 and 3 are similar to those of DGP 1, as shown in Section S.5 in the Supplementary Materials.

\begin{remark}
Additionally, we compare the results of model averaging by LOOCV (or jackknife model averaging, JMA) and JCVMA. We consider DGP 1 in Section \ref{subsec:sd1} as an example. For $n = 400, p=1,2,$ and $\tau=0.5,0.05$, we compare EFPE of JCVMA5 and JMA, as well as their computing time. As shown in Figure S.9 in the Supplementary Materials, the EFPEs of the two methods are similar. However, for $p=1$, the calculation time of JMA is approximately 65 times that of JCVMA5. For $p=2$, the calculation time of JMA is approximately 10 times that of JCVMA5.
\end{remark}

\subsubsection{Simulation Design II.}
In Simulation Design II, we assume that the data are generated from one of the candidate models,
i.e., the set of candidate models includes at least one correct model.
We consider the following DGP: 
$$y_{i}= \sum_{j=1}^{5}\beta_j x_{i j}+\epsilon_{i},$$
where $x_{i 1}=1$, $x_{i j}, j=2,3, \ldots,5$ are IID $N(0,1)$ and mutually independent of each other, 
$(\beta_1,\beta_2,\ldots,\beta_{5})=(1,1,1,1,0) $,
and $\epsilon_i$ is $N(0,1)$ and independent of $x_{i j}, j=2,3, \ldots,5$. Assume that the first variable $x_{i 1}$ must be included in the candidate models and the last four uncertain variables may not be included in the candidate models. Consider non-nested candidate models, i.e., $M=2^4=16$, where there are two correct models among the candidate models.

For the $r$th replication, the mean squared error (MSE) is calculated as follows:
$$\mathrm{MSE}(r)=\left\|\hat{\boldsymbol{\Theta}}^{(r)}\left(\hat{\mathbf{w}}\right)-\boldsymbol{\Theta}_{0}\right\|^2 ,
$$ 
where 
$\hat{\boldsymbol{\Theta}}^{(r)}\left(\hat{\mathbf{w}}\right)$ denotes the model averaging estimator of 
$\boldsymbol{\Theta}_{0}$.
We then average the MSE over $R = 500$ replications and obtain:
$
\mathrm{MSE}=\frac{1}{R} \sum_{r=1}^{R} \mathrm{MSE}(r).
$
Similarly, we calculate the average sum of weights $\hat{\tau}$ given to the correct models based on 500 replications.

We consider the cases $n=100, 200, 400, 800, 1600$, $\tau =0.5, 0.05$, and $p=1, 2$ with both JCVMA5 and JCVMA10. The results are shown in Tables \ref{table:MSE} and \ref{table:weight}, respectively. As $n$ increases, the MSE decreases to 0, demonstrating the consistency of  $\hat{\boldsymbol{\Theta}}\left(\hat{\mathbf{w}}\right)$. Moreover, the sum of the weights given to the correct models approaches 1 as $n$ increases, which demonstrates that $\hat{\tau}$ converges to 1 in probability.

\begin{table}[htbp]
\centering
\caption{MSE of the model averaging estimators}
\begin{tabular}{ccccccc}
\hline
& $n$       & 100   & 200   & 400   & 800   & 1600  \\ \hline
\multirow{2}{*}{$p=1, \tau=0.5$}  & JCVMA5  & 0.091 &	0.042 &	0.019 &	0.010 &	0.004  \\
   & JCVMA10 & 0.090 &	0.042 &	0.018 &	0.017 &	0.004  \\ 
\multirow{2}{*}{$p=1, \tau=0.05$} & JCVMA5  & 0.335 &	0.150 &	0.070 &	0.029 &	0.015  \\
   & JCVMA10 & 0.333 &	0.147 &	0.066 &	0.029 &	0.014  \\ 
\multirow{2}{*}{$p=2, \tau=0.5$}  & JCVMA5  & 0.050 &	0.023 &	0.011 &	0.005 &	0.003  \\
   & JCVMA10 & 0.049 &	0.023 &	0.010 &	0.005 &	0.003  \\ 
\multirow{2}{*}{$p=2, \tau=0.05$} & JCVMA5  & 0.147 &	0.067 	&0.036 &	0.020 &	0.012  \\
  & JCVMA10 & 0.144 &	0.066 &	0.036 &	0.020 &	0.012   \\ \hline
\end{tabular}
\label{table:MSE}
\end{table}

\begin{table}[htbp] 
\centering
\caption{The sum of weights given to the correct models}
\begin{tabular}{ccccccc}
\hline
           & $n$       & 100   & 200   & 400   & 800   & 1600  \\ \hline
\multirow{2}{*}{$p=1, \tau=0.5$}  & JCVMA5  & 0.842 &	0.913 &	0.956 &	0.977 &	0.987  \\
    & JCVMA10 & 0.850 &	0.923 &	0.959 &	0.978 &	0.989  \\ 
\multirow{2}{*}{$p=1, \tau=0.05$} & JCVMA5  & 0.668 &	0.806 &	0.891 &	0.945 &	0.969  \\
    & JCVMA10 & 0.666 &	0.816 &	0.901 &	0.948 &	0.972   \\ 
\multirow{2}{*}{$p=2, \tau=0.5$}  & JCVMA5  & 0.947 &	0.973 	& 0.987 &	0.994 &	0.997  \\
    & JCVMA10 & 0.958 &	0.980 &	0.990 &	0.995 &	0.998   \\ 
\multirow{2}{*}{$p=2, \tau=0.05$} & JCVMA5  & 0.800 &	0.889 &	0.942 &	0.969 &	0.981  \\
    & JCVMA10 & 0.810 &	0.901 &	0.950 &	0.973 &	0.984    \\ \hline
\end{tabular}
\label{table:weight}
\end{table}

\section{Empirical Evaluation}
\label{sec:emprical}
In this section, we validate the theoretical results by applying JCVMA to predict with a real data set: the collection volume of shipping carrier stores.

\subsection{Prediction of the Collection Volume of Shipping Carrier Stores}
Our objective is to predict the collection volume of shipping carrier stores. A proper prediction of the collection volume of shipping carrier stores will help improve operations management, thus creating improved customer service and a significant decrease in waste.
Note that an overestimation of collection volume may lead to idle resources. Meanwhile, an underestimation can lead to untimely delivery, thus affecting service quality. Therefore, a flexible loss function is needed. 

We use the collection volume of the shipping carrier stores and other related variables with a sample size of $n=175$.
The logarithm of the monthly collection volume of different outlets is the response variable, and there are 14 potential predictors. Since it is time-consuming to consider all potential models, we first sequence the variables and then use nested models as candidate models. 
For candidate models, we consider two different designs. In Design I, the variables are sorted by the absolute value of the correlation coefficient with the response variable, while nested candidate models are used. In Design II, we first conduct the least squares regression of the full model, then sort the variables according to the $p$-value corresponding to the significance test of each variable and use nested candidate models.

We compare the out-of-sample prediction performance of JCVMA with alternative model averaging methods, such as SAIC, SBIC, and EWA.  Because JCVMA5 is almost indistinguishable from JCVMA10 and model averaging outperforms model selection, we consider the following four model averaging methods: JCVMA5,  SAIC, SBIC, and EWA. 

\subsection{Prediction Results in Two Designs}
\subsubsection{Design I.}

In Design I, the variables are sorted by the absolute value of the correlation coefficient with the response variable, as shown in \cref{table:correlation}.
We consider the following 15 nested candidate models: $\{1\},\left\{1, x_{1}\right\}, \ldots, \left\{1, x_{1}, x_{2}, \ldots, x_{14}\right\}$.
\begin{table}[htbp]
\centering
\caption{Covariates and correlation}
\begin{tabular}{lll}
\hline
Variable & Name        & Correlation \\ \hline
$x_{1}$    & The logarithm of delivery volume       & \ 0.531       \\
$x_{2}$    & False sign-off rate          & -0.410      \\
$x_{3}$     & Loss rate        & -0.390      \\
$x_{4}$     & Total complaint rate      & -0.383      \\
$x_{5}$     & Same-day sign-off rate    & \ 0.362       \\
$x_{6}$     & On-time delivery rate      & \ 0.337       \\
$x_{7}$    & Secondary complaint rate        & -0.300      \\
$x_{8}$    & One-time resolution rate     & \ 0.274       \\
$x_{9}$   & Escalation complaint rate      & -0.250      \\
$x_{10}$    & Timely collection rate         & \ 0.238       \\
$x_{11}$    & Outlet score        & \ 0.123       \\
$x_{12}$    & Signed positive feedback rate    & \ 0.079       \\
$x_{13}$    & Timely departure rate      & -0.063      \\
$x_{14}$    & Positive feedback from the representative signatory & \ 0.055  \\ \hline    
\end{tabular}
\label{table:correlation}
\end{table}

We randomly divide the samples into a training set containing $n_1$ samples and a validation set $\left\{(y_{s}, \mathbf{x}_{s})\right\}_{s=1}^{n_s}$, where $\mathbf{x}_{s}=(1,x_{s1},\ldots , x_{s 14})$, and $n_s=n-n_1$. 
Since we do not know the true DGP, we can only evaluate the predicting performance of JCVMA.
We use the FPE to evaluate the prediction of sales; that is:
$$
\operatorname{FPE}=\frac{1}{n_s} \sum_{s=1}^{n_s} \rho_{\tau,p}\left(y_{n_s}-\hat{\mu}_{n_s} \right),
$$
where $\hat{\mu}_{n_s}$ is estimated using each method. The FPE is averaged after repeating $R=500$ times. 

For easy comparison, we normalize the FPE by dividing it by the FPE of the largest model and then reporting the relative FPE. A lower relative FPE is indicative of better predictive performance. When the relative FPE exceeds 1, this indicates that the specified method performs worse than the largest model. In this case, instead of using the new method, it is better to simply use the largest model.

We consider the cases $n_1=50, 100, 150$, $\tau=0.95, 0.05$, and $p=1, 2$. The relative FPEs are shown in \cref{table:FPE}.
For clarity, the two best performing methods are labeled as [1] and [2]  in the upper right corner, respectively.

\begin{table}[htbp]
\centering
\caption{Relative FPE of the prediction of collection volume in Design I}
\begin{tabular}{ccccccc}
\hline
$p$ & $\tau$  & $n_1$   & JCVMA5  & SAIC  & SBIC  & EWA    \\ \hline
1 & 0.5  & 50  & 0.529$^{[1]}$    & 0.943 & 0.639$^{[2]}$ & 0.676  \\
1 & 0.5  & 100 & 0.641$^{[1]}$    & 0.991 & 0.700$^{[2]}$ & 0.759  \\
1 & 0.5  & 150 & 0.753$^{[1]}$    & 0.968 & 0.782$^{[2]}$ & 0.823  \\
1 & 0.05 & 50  & 0.340$^{[1]}$   & 0.936 & 0.858 & 0.509$^{[2]}$  \\
1 & 0.05 & 100 & 0.302$^{[1]}$   & 0.972 & 0.921 & 0.376$^{[2]}$  \\
1 & 0.05 & 150 & 0.392$^{[2]}$    & 0.952 & 0.888 & 0.319$^{[1]}$ \\
2 & 0.5  & 50  & 0.231$^{[1]}$    & 0.931 & 0.286$^{[2]}$ & 0.370  \\
2 & 0.5  & 100 & 0.283$^{[1]}$    & 0.959 & 0.305$^{[2]}$ & 0.437 \\
2 & 0.5  & 150 & 0.248$^{[1]}$     & 1.014 & 0.419$^{[2]}$ & 0.711  \\
2 & 0.05 & 50  & 0.144$^{[1]}$    & 0.947 & 0.635 & 0.350$^{[2]}$  \\
2 & 0.05 & 100 & 0.069$^{[1]}$   & 0.871 & 0.451 & 0.122$^{[2]}$ \\
2 & 0.05 & 150 & 0.070$^{[2]}$   & 0.864 & 0.530 & 0.037$^{[1]}$ \\ \hline
\end{tabular}
\label{table:FPE}
\end{table}

Interestingly, JCVMA5 is always one of the two best performing methods, exhibiting the best performance in most cases. EWA and SBIC sometimes perform well and sometimes poorly. SAIC  always performs poorly, especially when $p=2, \tau=0.5, n_1=150$, while the relative FPE is greater than 1, which means the model averaging estimator of SAIC is poorer than the largest model. 
Importantly, JCVMA is significantly better than the largest model in all cases and the advantages of asymmetrical situations are more obvious. 
When $p=1$, $\tau=0.5$, for different settings of $n$, the FPE of JCVMA5 is 53--75\% of the FPE of the largest model. When $p=1$, $\tau=0.05$, the FPE of JCVMA5 is only about 30\% of the FPE of the largest model.
Besides, when $p=2$, $\tau=0.5$, the FPE of JCVMA5 is about 25\% of the FPE of the largest model.
When $p=2$, $\tau=0.05$, the FPE of JCVMA5 is only about 10\% of the FPE of the largest model, which means that the proposed model averaging methods has an enormous improvement on prediction.

\subsubsection{Design II.}
In Design II, we first conduct the least squares regression of the full model and then sort the variables according to the $p$-value corresponding to the significance test of each variable, as shown in \cref{table:p_value}. Then, we consider the following 15 nested candidate models: $\{1\},\left\{1, x_{1}\right\}, \ldots, \left\{1, x_{1}, x_{2}, \ldots, x_{14}\right\}$.

\begin{table}[htbp]
\centering
\caption{Covariates and $p$-value}
\begin{tabular}{lll}
\hline
Variable & Name      & $p-$value \\ \hline
$x_1$        & Delivery volume       & 0.0000  \\
$x_2$        & Timely collection rate      & 0.0001  \\
$x_3$        & Outlet score    & 0.0001  \\
$x_4$        & One-time resolution rate   & 0.0001  \\
$x_5$        & On-time delivery rate   & 0.0017  \\
$x_6$        & Positive feedback from the representative signatory & 0.2531  \\
$x_7$        & Signed positive feedback rate     & 0.2641  \\
$x_8$        & Secondary complaint rate      & 0.4098  \\
$x_9$        & False sign-off rate  & 0.4552  \\
$x_{10}$       & Escalation complaint rate     & 0.5475  \\
$x_{11}$       & Timely departure rate     & 0.7401  \\
$x_{12}$       & Same-day sign-off rate      & 0.7929  \\
$x_{13}$       & Total complaint rate     & 0.8499  \\
$x_{14}$       & Loss rate      & 0.9842 \\ \hline
\end{tabular}
\label{table:p_value}
\end{table}

Similar to Design I, we normalize FPE by dividing it by the FPE of the largest model and report the relative FPE. 
The relative FPEs are visualized in \cref{table:FPE_2}.

\begin{table}[htbp]
\centering
\caption{Relative FPE of the prediction of collection volume in Design II}
\begin{tabular}{ccccccc}
\hline
$p$ & $\tau$  & $n_1$   & JCVMA5  & SAIC  & SBIC  & EWA    \\ \hline
1 & 0.5  & 50  & 0.305$^{[2]}$   & 0.647 & 0.316  & 0.305$^{[1]}$  \\
1 & 0.5  & 100 & 0.507$^{[1]}$   & 0.648 & 0.509 & 0.508$^{[2]}$  \\
1 & 0.5  & 150 & 0.695  & 0.762 & 0.677$^{[1]}$  & 0.686$^{[2]}$  \\
1 & 0.05 & 50  & 0.152$^{[1]}$   & 0.627 & 0.404 & 0.196$^{[2]}$  \\
1 & 0.05 & 100 & 0.201$^{[1]}$   & 0.670 & 0.386 & 0.238$^{[2]}$  \\
1 & 0.05 & 150 & 0.233$^{[1]}$   & 0.529 & 0.322 & 0.267$^{[2]}$  \\
2 & 0.5  & 50  & 0.206$^{[1]}$   & 0.932 & 0.330$^{[2]}$  & 0.352 \\
2 & 0.5  & 100 & 0.239$^{[1]}$   & 0.990 & 0.309$^{[2]}$  & 0.441 \\
2 & 0.5  & 150 & 0.394$^{[1]}$   & 1.003 & 0.463$^{[2]}$  & 0.554 \\
2 & 0.05 & 50  & 0.106$^{[1]}$   & 0.933 & 0.526 & 0.351$^{[2]}$  \\
2 & 0.05 & 100 & 0.074$^{[1]}$   & 0.926 & 0.520 & 0.130$^{[2]}$  \\
2 & 0.05 & 150 & 0.071$^{[2]}$   & 0.865 & 0.681 & 0.055$^{[1]}$  \\ \hline
\end{tabular}
\label{table:FPE_2}
\end{table}

 The result is similar to that of Design I.
We can see that JCVMA5 performs the best in most cases. EWA and SBIC sometimes perform well and sometimes poorly. SAIC  always performs poorly, especially when $p=2$,$\tau=0.5$,$n_1=150$, the relative FPE is greater than 1, which means the model averaging estimator of SAIC is poorer than the largest model. 
Moreover, JCVMA is significantly better than the largest model in all cases and the advantages are more obvious than Design I. Except for the case when $p=1$, $\tau=0.5$, the FPE of JCVMA5 is less than 40\% of the FPE of the largest model. 
Besides, when $\tau=0.05$, the FPE of JCVMA5 is less than 25\% of the FPE of the largest model, which means that the improvement of JCVMA for asymmetric loss is larger than for symmetric loss.

Additionally, we apply our JCVMA to forecast excess stock returns. This analysis can be found in Section S.2 of the Supplementary Materials, which further illustrates the superiority of our method.

\section{Conclusions}
\label{sec:conclusions}
We propose a model averaging method under a flexible loss function with a J-fold cross-validation criterion to determine the weights. Both theoretical findings and simulation results guarantee the superiority of the proposed method. We apply our method to predict the collection volume of shipping carrier stores and excess stock returns under a flexible loss function, demonstrating our method's advantages.

Several relevant questions deserve future research. First, although we have proved the convergence of the model averaging estimator, the convergence rate is slower than that of the correct models; we could develop new model averaging methods to ensure a faster convergence rate. Second, we can extend our results to more general loss functions, such as the Bregman loss \citep{bregman1967relaxation}. Third, developing a model averaging method under a flexible loss function for more general models would be interesting.  For instance, we could accommodate the nonlinearity of covariates, dependent data, and high-dimensional models.

\bibliographystyle{apalike}
\bibliography{reference}

\end{document}